\documentclass[aps,prb,twocolumn,amsmath,amssymb,groupedaddress,showpacs,10pt,superscriptaddress,floatfix]{revtex4-1}

\usepackage[utf8]{inputenc}

\usepackage{graphicx}
\usepackage{dcolumn}
\usepackage{bm}
\usepackage[colorlinks=true,linkcolor=blue,citecolor=blue,urlcolor=blue]{hyperref}
\usepackage{color}
\usepackage{amsmath}
\usepackage{siunitx}
\usepackage{dcolumn}

\begin{document}
\newcommand{\nvm}{$\mathrm{NV^{\mbox{-}}}\ $}
\newcommand{\nvmnosp}{$\mathrm{NV^{\mbox{-}}}$}
\newcommand{\nvz}{$\mathrm{NV^0}\ $} 
\newcommand{\nvznosp}{$\mathrm{NV^0}$}
\newcommand{\nvratio}{$\mathrm{NV^{\mbox{-}}:NV^0}\ $} 
\newcommand{\mszero}{$\mathrm{m_S = 0}$}
\newcommand{\msone}{$\mathrm{m_S = \pm 1}$} 

\title{Creation of ensembles of nitrogen-vacancy centers in diamond by neutron and electron irradiation}


\author{Tobias N\"{o}bauer}
\email[]{tobias.noebauer@tuwien.ac.at}
\author{Kathrin Buczak}
\author{Andreas Angerer}
\author{Stefan Putz}
\affiliation{Vienna Center for Quantum Science and Technology and Atominstitut, TU Vienna, Stadionallee 2, 1020 Vienna, Austria, EU}
\author{Georg Steinhauser}
\affiliation{Department of Environmental and Radiological Health Sciences, Colorado State University, Fort Collins, CO 80523-1618, United States}
\author{Johanna Akbarzadeh}
\author{Herwig Peterlik}
\affiliation{Dynamics of Condensed Systems group, Faculty of Physics, University of Vienna, Boltzmanngasse 5, 1090 Vienna, Austria, EU}
\author{Johannes Majer}
\author{J\"{o}rg Schmiedmayer}
\author{Michael Trupke}
\affiliation{Vienna Center for Quantum Science and Technology and Atominstitut, TU Vienna, Stadionallee 2, 1020 Vienna, Austria, EU}

\date{\today}

\begin{abstract}
We created dense ensembles of negatively charged nitrogen-vacancy (\nvmnosp) centers in diamond by neutron and electron irradiation for applications in hybrid quantum systems and magnetometry. We characterize fluorescence intensity, optical and coherence properties of the resulting defects by confocal microscopy, UV/Vis and FTIR spectroscopy, optically detected magnetic resonance and small angle X-ray scattering. We find the highest densities of \nvm at neutron fluences on the order of \SI{e17}{\per \square \cm} and electron doses of \SI{e18}{\per\square\cm}, with spin resonance linewidths of \SI{6}{MHz}. Lower electron energies increase the ratio of centers in the desired negative charge state to those in the neutral one. Annealing at \SI{900}{\celsius} during the irradiation reduces the spin resonance linewidth. Electron irradiation furthermore results in substantially higher optical transparency compared to neutron irradiation.
\end{abstract}


\maketitle

\section{Introduction}
Recently, ensembles of spins in diamond have shown great potential for practical quantum technologies: They have been coupled coherently to superconducting microwave resonators\cite{schuster_high-cooperativity_2010,kubo_strong_2010,amsuss_cavity_2011} with the perspective of employing the resulting hybrid quantum system for storing quantum information\cite{kubo_storage_2012,julsgaard_quantum_2013} in collective excitations. Spin ensembles have also been coupled to superconducting qubits\cite{zhu_coherent_2011,kubo_hybrid_2011}, which allow the creation and processing of single excitations in the microwave regime.

In addition to these quantum information technologies, especially negatively charged nitrogen-vacancy (\nvmnosp) center spins in diamond show great promise for ultra-sensitive magnetometry, based on both single spins\cite{maze_nanoscale_2008,balasubramanian_nanoscale_2008} and spin ensembles\cite{taylor_high-sensitivity_2008,pham_magnetic_2011,acosta_broadband_2010}. In contrast to existing technologies such as atomic vapour cells or SQUIDs, \nvmnosp-based magnetometry schemes combine small probe volumes with high sensitivities, and can be operated at temperatures ranging from a few Kelvin to above room temperature. This is made possible by remarkably high spin coherence times of up to several ms in isotopically purified samples\cite{balasubramanian_ultralong_2009}. A convenient level scheme involving a spin-selective shelving process allows the Zeeman shift in the spin triplet ground state of \nvm centers to be read out optically.

A number of approaches has been evaluated which have the aim of creating dense ensembles of \nvm while maintaining a low inhomogeneous spin resonance linewidth, as is required both for hybrid quantum systems as well as ensemble-based metrology\cite{acosta_diamonds_2009,aharonovich_producing_2009,botsoa_optimal_2011,schwartz_effects_2012,kim_electron_2012}. In the present study we use neutron and electron irradiation under various conditions to find optimized parameters for the creation of \nvm defects. We characterize the resulting fluorescence intensities and spectra (Sec.~\ref{sec_fluo}), visual and infrared absorption properties (Sec.~\ref{sec_vis},~\ref{sec_ir}) as well as spin resonance parameters (Sec.~\ref{sec_odmr}). We contrast the optical measurements with results from small angle X-ray scattering analysis (Sec.~\ref{sec_saxs}). In Sec.~\ref{sec_discussion} we discuss our results with regard to the aforementioned application areas.

\section{Experimental Procedure}
\subsection{Raw samples}
Three classes of samples were treated and examined in the course of this work: Type IIa single crystal diamonds produced from a chemical vapour deposition (CVD) process by Element-6, Type Ib single crystals synthesized using the high-pressure-high-temperature (HPHT) method by Element-6, and similar Type Ib samples by Sumitomo. All samples are millimetre-sized and consist to a large extent of a single growth sector (Element-6 specifies typically \textless 70-80\%). The concentration of nitrogen ranges from \text{1-\SI{200}{ppm}} (where \SI{1}{ppm} corresponds to \SI{1.76e17}{\per\cubic\cm} in diamond). The next most abundant impurity element in all Element-6 samples is boron, at concentrations of \textless \SI{0.1}{ppm} and \textless \SI{0.05}{ppm} in the HPHT and CVD samples, respectively (no data is available for non-nitrogen impurities in the Sumitomo samples). The initial concentration of NV centers is \SI{\approx 1e17}{\per\cubic\cm} in the HPHT and \SI{\approx 4e14}{\per\cubic\cm} in the CVD samples, respectively\footnote{Private communication with Element-6.}. A summary of sample parameters and some key results is given in Table~\ref{tab:data}.

\begin{table*}
\caption{\label{tab:data}Raw sample parameters and main results. In column ``Irradiation'', n and e indicate neutron and electron irradiation, respectively. For electron-irradiated samples, electron energy and temperature during irradiation are also given in this column.}
\begin{ruledtabular}
\begin{tabular}{ l c c c c r c c }
  Sample		 	& 	Company		& 		Synthesis	& initial [N$\mathrm{_S}$] & 	Irradiation			& Fluence			& final [NV$\mathrm{^{\mbox{-}}}]$ estimate 	& final $\mathrm{T_2^*}$	\\
   	& 	& 	& [ppm] &  	& [\si{\per \square \cm}]	& [a.u.]		 	& [ns]\\
  
  \hline
  L1  		& E6			& HPHT		& $\lesssim 200$			& n				& \num{2.5(5)e17}		& \num{323(36)}				& \num{46.7(4)}						\\
  E1  		& E6			& HPHT		& $\lesssim 200$			& n				& \num{4.1(8)e17}		& \num{402(23)}				& \num{54.8(12)}						\\
  H4b  		& E6			& HPHT		& $\lesssim 200$			& n				& \num{5.4(11)e17}		& \num{435(34)}				& \num{60.3(22)}						\\
  BS3-1b  	& E6			& HPHT		& $\lesssim 200$			& n				& \num{9.0(18)e17}		& \num{468(59)}				& \num{50.8(5)}						\\
  BS3-2a  	& E6			& HPHT		& $\lesssim 200$			& n				& \num{1.4(2)e18}		& \num{390(250)}				& \num{42}						\\
  BS3-3a  	& E6			& HPHT		& $\lesssim 200$			& n				& \num{1.8(3)e18}		& \num{357(87)}				& \num{40.9(19)}						\\
  BS3-3b  	& E6			& HPHT		& $\lesssim 200$			& n				& \num{1.8(3)e18}		& \num{294(10)}				& \num{41.8(18)}						\\
  BS3-4a  	& E6			& HPHT		& $\lesssim 200$			& n				& \num{2.2(4)e18}		& \num{290(230)}				& \num{32.2(10)}						\\
  BS3-4b  	& E6			& HPHT		& $\lesssim 200$			& n				& \num{2.2(4)e18}		& \num{161(4)}				& \num{44.9(25)}						\\
  BS3-5a  	& E6			& HPHT		& $\lesssim 200$			& n				& \num{2.7(5)e18}		& \num{230(79)}				& \num{48.4(12)}						\\
  BS3-5b  	& E6			& HPHT		& $\lesssim 200$			& n				& \num{2.7(5)e18}		& \num{262(51)}				& \num{29.4(8)}						\\
  \hline
  S1  		& E6			& HPHT		& $\lesssim 200$			& e, \SI{8}{\MeV}, \SI{100}{\celsius}		& \num{1.59e17}			& 				& 					\\
  Q4  		& E6			& HPHT		& $\lesssim 200$			& e, \SI{8}{\MeV}, \SI{100}{\celsius}		& \num{7.80e18}			& 				& 						\\
  Q5  		& E6			& HPHT		& $\lesssim 200$			& e, \SI{8}{\MeV}, \SI{100}{\celsius}		& \num{1.57e19}			& 				& 						\\
  Q6  		& E6			& HPHT		& $\lesssim 200$			& e, \SI{8}{\MeV}, \SI{100}{\celsius}		& \num{1.57e19}			& 				& 						\\
  \hline
  U5  		& E6			& HPHT		& $\lesssim 200$			& e, \SI{6.5}{\MeV}, 750-\SI{900}{\celsius}		& \num{1.0e18}			& 		\num{242(227)}		& \num{58}	\\
  U6  		& E6			& HPHT		& $\lesssim 200$			& e, \SI{6.5}{\MeV}, 750-\SI{900}{\celsius}		& \num{1.0e18}		    & \num{83(10)}	& \num{49}	\\
  U7  		& E6			& HPHT		& $\lesssim 200$			& e, \SI{6.5}{\MeV}, 750-\SI{900}{\celsius}		& \num{1.0e18}			& \num{154(25)}				& \num{49}	\\
  \hline
  U9  		& E6			& HPHT		& $\lesssim 200$			& e, \SI{6.5}{\MeV}, \SI{100}{\celsius}		& \num{1.0e18}			& \num{93(51)}					& \num{27}	\\
  U10  		& E6			& HPHT		& $\lesssim 200$			& e, \SI{6.5}{\MeV}, \SI{100}{\celsius}		& \num{1.0e18}			& \num{196(29)}				& \num{40}	\\

  \hline
  C5  		& E6			& HPHT		& $\lesssim 200$			& -											& \num{0}				& 				& 	\\	
  \hline				
  J6  		& Sumitomo	& HPHT		& $\approx 100$			& n				& \num{5.4(11)e17}		& 273.2(1)				& \num{119(18)}						\\
  G3  		& Sumitomo	& HPHT		& $\approx 100$			& n				& \num{5.4(11)e17}		& 				& 						\\
  \hline
  D2  		& E6			& CVD		& $\lesssim 1$			& n											& \num{5.2 +- 1e17}		& \num{<0.01} 				& \num{171(13)}	\\
  Y1  		& E6			& CVD		& $\lesssim 1$			& e, \SI{6.5}{\MeV}, 750-\SI{900}{\celsius}		& \num{1.0e16}			& 				& \num{350}	\\
  Y4  		& E6			& CVD		& $\lesssim 1$			& e, \SI{6.5}{\MeV}, 750-\SI{900}{\celsius}		& \num{1.0e17}			& \num{<0.1}				& \num{371}	\\
  Y7  		& E6			& CVD		& $\lesssim 1$			& e, \SI{6.5}{\MeV}, 750-\SI{900}{\celsius}		& \num{1.0e18}			& \num{<0.1}				& \num{304}	\\
\end{tabular}

\end{ruledtabular}
\end{table*}

\subsection{Irradiation and annealing}
\subsubsection{Neutron irradiation}
A number of samples were subject to irradiation in the core of the \SI{250}{\kilo\watt_{th}} General Atomics TRIGA Mark~II nuclear reactor at Atominstitut, TU Vienna. The samples were placed inside a sealed quartz tube and a graphite irradiation capsule in the center of the reactor core, where a total neutron flux density of \SI{5 +- 1 e12}{\per\square\cm \per\second} was maintained. Neutrons with energies lower than \SI{\approx 100}{\electronvolt} cannot transfer the displacement energy to carbon in the diamond lattice (\SI{32}{\electronvolt}\cite{davies_chapter_1994}) in an elastic scattering event and do not contribute to vacancy creation, reducing the effective flux density by \SI{\approx 30}{\percent}. The remaining neutron energies are distributed evenly in the energy range up to \SI{1e5}{\electronvolt}, followed by a peak in flux density around \SI{2.5}{\mega \electronvolt} and a sharp cut-off above this energy\citep{weber_neutron_1986}. 

Irradiation with a neutron fluence $F$ results in $N=F n_0 \sigma_{el}$ knock-on collisions, where $n_0=\SI{1.76e23}{\per\cubic\cm}$ is the atom number density of diamond, and $\sigma_{el}$ the elastic neutron scattering cross section. For neutrons in the relevant energy range from \SI{1}{\kilo \electronvolt} to \SI{1}{\mega \electronvolt}, $\sigma_{el}$ \SIrange{\approx 4.8}{2.4}{\barn}\cite{chadwick_endf/b-vii.1_2011}, resulting in $N\approx 0.7 F$, i.e. nearly one primary knock-on per incident neutron~\cite{morelli_phonon_1993}. An average energy of \SI{140}{\electronvolt} to \SI{140}{\kilo \electronvolt} is transferred to a carbon recoil nucleus by neutrons in this energy range~\cite{seitz_disordering_1949}, causing a substantial recoil which displaces further carbon atoms and creates a region of disordered carbon roughly \SI{45}{\angstrom} in diameter~\cite{dienes_nature_1953}.

Based on previous work~\cite{mita_change_1996}, we chose to test irradiation fluences in the range \SIrange{1.2e17}{1.3e18}{\per \square \cm} in order to optimize \nvm production, requiring \SIrange{14}{150}{\hour} of irradiation. Temperatures during irradiation did not exceed \SI{80}{\celsius}. The cross sections for non-elastic nuclear reactions such as neutron capture are at least four orders of magnitude smaller than the elastic scattering cross section at all relevant energies and do not significantly alter the isotope composition of the crystal.

\subsubsection{Electron irradiation}
For comparison with previously reported results\cite{acosta_diamonds_2009}, we also exposed HPHT and CVD diamond samples to electron doses ranging from \SIrange{1e16}{2e19}{\per \square \cm} at the linear accelerator of the Istituto per la Sintesi Organica e la Fotoreattivit\`{a} in Bologna, Italy. Either \SI{6.5}{\mega \electronvolt} or \SI{8}{\mega \electronvolt} electrons were delivered in pulses of \SI{5}{\micro \second} at a repetition rate of \SI{50}{\Hz} and a peak pulse current of \SI{0.8}{\A}, yielding an average current of \SI{150}{\micro \ampere}. While some samples were maintained at low temperatures (\SI{<100}{\celsius}), others where heated to \SIrange{750}{900}{\celsius} during irradiation in order to alter vacancy mobility and recombination dynamics, with the aim of preventing the formation of defect clusters.

A simulation model presented in Ref.~\onlinecite{campbell_radiation_2000} yields a rate of vacancy production from electron irradiation in diamond of 2.85 (3.42)\,vacancies/e$^-$/cm for 5 (10)\,\si{\mega \electronvolt} electrons, respectively. The model seems to slightly over-estimate the rates for lower energy electrons compared to experimental findings\cite{hunt_identification_2000,newton_recombination-enhanced_2002}, so we assume a rate of $\approx 2.7\,$(2.8) vacancies/e$^-$/cm at \SI{6.5}{\mega \electronvolt} (\SI{8}{\mega \electronvolt}), respectively. Since the stopping range of electrons in diamond (\SI{1.1}{\cm} at \SI{6.5}{\mega \electronvolt} and  \SI{1.3}{\cm} at \SI{8}{\mega \electronvolt} in the continuous slowing down approximation\cite{berger_estar_2013}) is much larger than the sample thickness, energy is deposited uniformly throughout the crystal. 

\subsubsection{Annealing}
All samples that were not kept at high temperatures during irradiation were subsequently annealed in an argon atmosphere at \SI{900}{\celsius} for three hours. This treatment allows the created vacancies to become mobile and pair with N impurities to form NV centers. Annealing at higher temperatures up to \SI{1050}{\celsius}, or for longer times, does not significantly improve NV yield\cite{acosta_diamonds_2009}.

\subsection{Material analysis}
\subsubsection{Confocal microscopy and spectrophotometry}
\label{subsubsec:confocal}
In order to characterize fluorescence intensities and spectra the annealed samples were examined in a home-built room-temperature confocal microscope. Light from a commercial diode-pumped solid state laser at \SI{532}{\nm} was combined into the collection beam path via reflection from an uncoated glass wedge and focused onto the samples by an Olympus PLAPON 60XO $\text{NA}=1.42$ oil immersion objective, exciting both \nvm and \nvz fluorescence (the latter at somewhat higher efficiency), which was collected by the same objective. Two detection modes were used in this work: 

For confocal microscopy and fluorescence intensity measurements, light from the sample was imaged onto a confocal pinhole by the tube lens, if necessary attenuated using calibrated neutral density filters, passed through a \SI{645}{\nm} long-wavelength-pass filter and imaged onto single photon counting modules. Some \SI{92}{\percent} of \nvm and \SI{29}{\percent} of \nvz fluorescence pass through this filter, resulting in an approximate overall photon detection efficiency of \SI{\sim 1}{\percent} for \nvmnosp. For spatial imaging, the samples were mounted onto a 3-axis piezo-actuated translation stage and scanned with respect to the objective. The confocal detection volume was determined to be \SI{\approx 2.3}{\cubic \micro \meter} by imaging a single \nvm center in ultrapure CVD diamond. From a single center our setup detects \SI{150000}{cps} near full saturation. 

For fluorescence spectrophotometry, light from the sample was reflected off a flip mirror inserted before the tube lens, passed through a \SI{532}{\nm} notch filter and coupled into a multi-mode fibre connected to an Ocean Optics Maya~2000~Pro spectrometer with an uncooled CCD detector. 

UV, visual and infrared absorption spectra were taken using a Perkin Elmer Lambda 750 UV/Vis and a Bruker Vector 22 FTIR spectrophotometer, respectively. 

\subsubsection{Optically detected magnetic resonance}
Due to spin-dependent intersystem crossing rates, \nvm centers are suitable for optically detected magnetic resonance (ODMR) techniques\cite{jelezko_single_2006} (cf. Section~\ref{sec_odmr} for details). We applied microwave fields via a coplanar waveguide structure on the sample holder and a \SI{100}{\micro \meter} gold wire placed across the sample. The MW frequency was scanned across the \nvm zero-field splitting frequency at \SI{2.87}{GHz} while exciting and counting fluorescence using the confocal microscope described in the previous section. ODMR spectra were recorded both in the earth's magnetic field and at higher fields created by a permanent magnet placed next to the samples (creating field strengths of hundreds of mT, not aligned with \nvm symmetry axes). The microwave power was scanned to provide data necessary to fit the power broadening and to extrapolate the bare linewidths.

\subsubsection{Small-angle X-ray scattering}
Selected samples were examined using Small-angle X-ray scattering (SAXS) in order to gain a deeper understanding of the induced lattice defects. These experiments were carried out with a rotating copper anode generator (Bruker Nanostar, superspeed solution), equipped with a 2D gas detector with microgap technology (Vantec 2000). The samples were measured for \SI{6}{\hour} perpendicular to the (100) surface of the diamond crystals. The sample-to-detector distance was \SI{108}{\cm}, giving access to a range of scattering vector magnitudes $q$ from \SI{0.1}{\per \nm} to \SI{2.8}{\per \nm}, where  $q=\frac{4\pi}{\lambda}\sin \theta$. Here, $2\theta$ is the scattering angle and $\lambda=\SI{0.1542}{\nm}$ the X-ray wavelength.

The obtained scattering patterns were radially averaged in order to obtain the scattered intensity in dependence of the magnitude of the scattering vector. For background correction an untreated diamond (Type Ib) was measured analogously to the other samples. The averaged scattering data were normalized to the diamond thickness using an X-ray transparent beamstop and corrected for background scattering by subtracting the untreated diamond sample.

\section{Results}
\subsection{Fluorescence}
\label{sec_fluo}

\begin{figure}
	\includegraphics[width=\columnwidth]{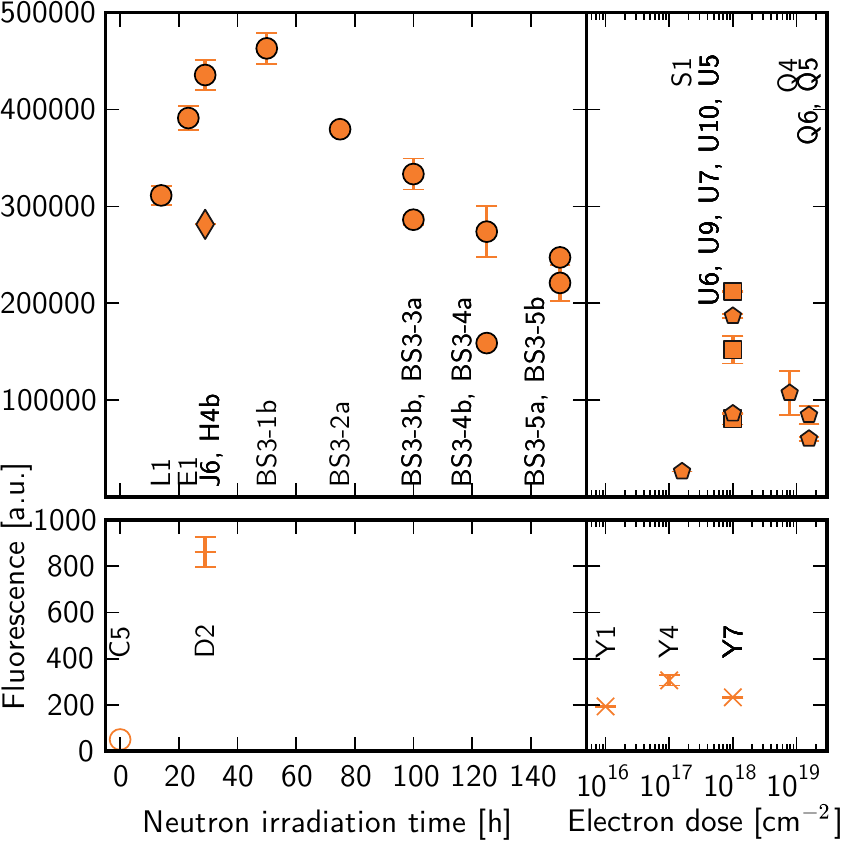}
	\caption{\label{fig_fluo}Fluorescence \SI{> 645}{\nm} vs. neutron irradiation time or electron dose. Error bars indicate standard errors. Upper panel: Type 1b HPHT samples. Lower panel: Type IIa CVD samples. Markers: $\bullet\ $Type 1b, neutrons. $\mbox{\tiny$\blacksquare$}\ $Type 1b, electrons, heated during irradiation. Pentagon:~Type 1b, electrons, RT during irradiation. $\circ\ $Type 1b, untreated, for comparison. $+\ $Type IIa CVD, neutrons. $\times\ $Type IIa CVD, electrons, heated during irradiation. $\diamond\ $Sumitomo Type 1b HPHT, neutrons.}
\end{figure}

Fluorescence depth profiles were taken at several locations within each sample by scanning the sample towards the objective while exciting with \SI{530}{\micro \watt} at \SI{532}{\nano \meter}. An exponential decrease of fluorescence with depth was observed in all samples, as expected from the absorption of both excitation and (to a lesser extent) fluorescence light. In the most strongly fluorescent samples, the recorded intensity decays to its $\frac{1}{e}$ value within a scanner displacement of \SI{11}{\micro \meter}; Note that this fluorescence attenuation coefficient does not directly correspond to the absorption coefficient of the samples as discussed in Sec.~\ref{sec_vis}. The absence of other variations along $z$ (at length scales larger than our axial resolution $\Delta z\approx \SI{3}{\micro \meter}$) indicates a uniform distribution of fluorescent defects, as expected from both electron and neutron irradiation.

\begin{figure}
	\includegraphics[width=\columnwidth]{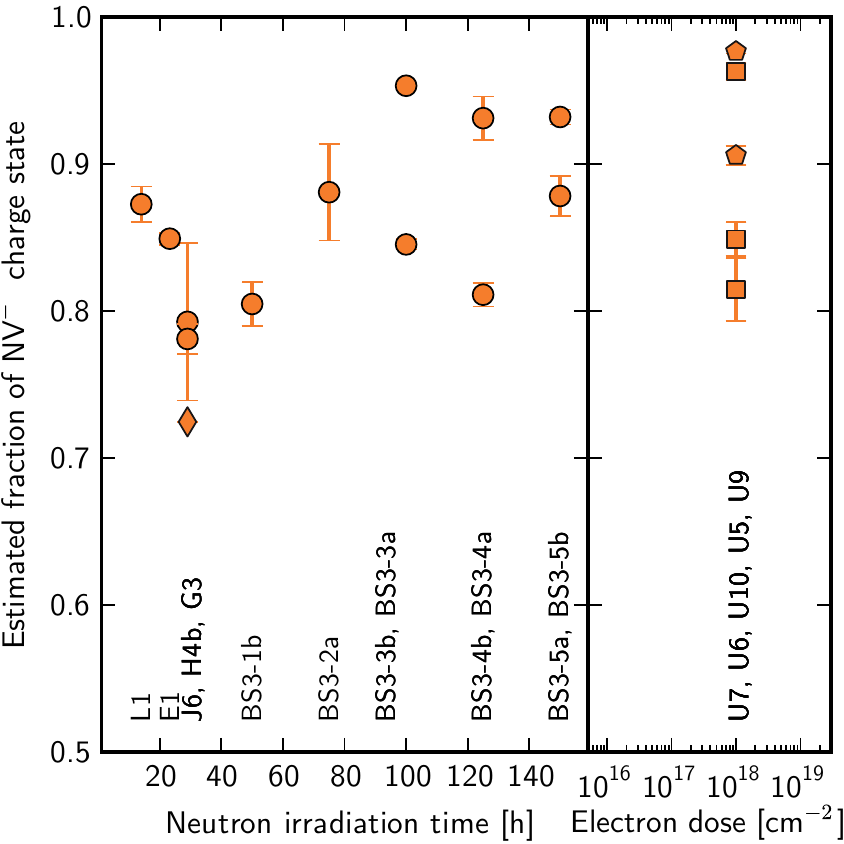}
	\caption{\label{fig_nvm_fraction}Estimated fraction of NV centers in the \nvm charge state vs. neutron irradiation time or electron dose. Error bars indicate standard errors. For marker legend, see caption of Fig.~\ref{fig_fluo}}
\end{figure}

In the fluorescence spectra recorded under the same excitation conditions (but with a different detection path, see Sec.~\ref{subsubsec:confocal}), the zero-phonon lines (ZPL) and phonon sidebands of both \nvz and \nvm defects can be identified at varying intensities. In order to compute an estimate of the fraction of centers in the \nvm charge state (under permanent \SI{532}{\nm} excitation), we first determine the background-subtracted intensities emitted into \SI{+- 6}{\nm} windows around the ZPLs of \nvm and \nvz at \SI{637}{\nm} and \SI{575}{\nm}, respectively. Due to the different Huang-Rhys factors \citep{acosta_diamonds_2009} as well as excitation cross sections and fluorescence dynamics of the two centers, the \nvratio concentration ratio is about a factor of four larger than the corresponding ZPL intensity ratio. After correcting for these effects, we obtain an estimate of the equilibrium ratio of NV centers in the \nvm charge state under permanent excitation which is shown in Fig.~\ref{fig_nvm_fraction}.

\begin{figure*}
	\includegraphics[width=\textwidth]{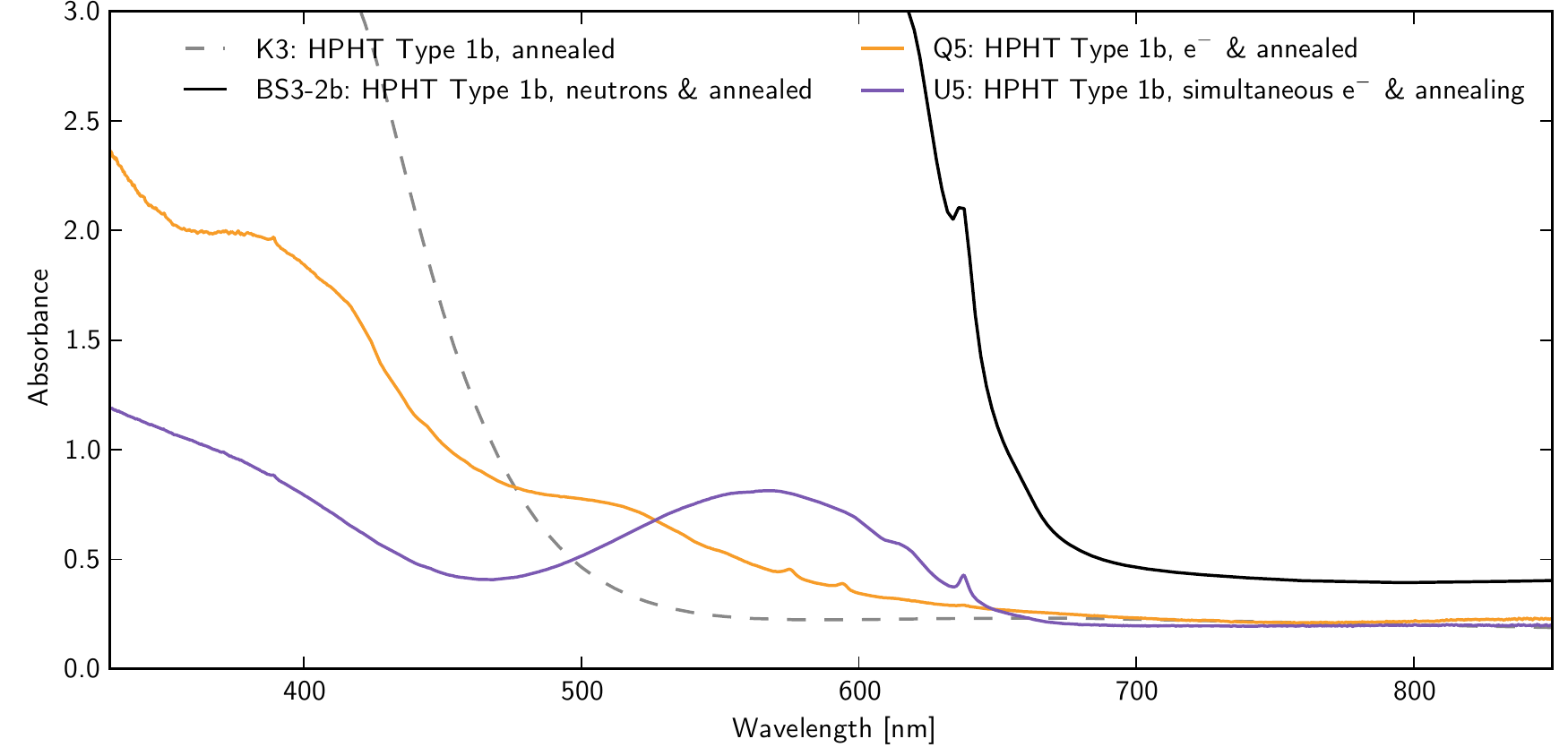}
	\caption{\label{fig_uvvis}(Color online) UV/Vis absorption spectra for four differently treated Type Ib HPHT samples. Sample thickness was \SI{0.5}{\mm} for samples K3 and BS3-3b, \SI{0.2}{\mm} for Q5 and \SI{0.1}{\mm} for U5.}
\end{figure*}
Within each sample, the maximum fluorescence intensities typically varied by \SI{+- 3}{\%} within the scan range of the piezo stage (\SI{20}{\micro \meter}), and by \SI{+- 15}{\%} across one sample (when crossing growth sector boundaries). The dependence of the averaged fluorescence intensity (filtered to \SI{> 645}{nm}) on the irradiation doses is shown in Fig.~\ref{fig_fluo}: for the neutron-irradiated samples with equal initial nitrogen concentrations (filled circles in upper left panel) it increases monotonically to a maximum at \SI{50}{\hour} of neutron irradiation, corresponding to a neutron fluence of \SI{\approx 4e17}{\per \square \centi \meter} in the energy range \SI{> 1}{\kilo \electronvolt}, which is in agreement with data published in Ref.~\onlinecite{mita_change_1996}. Sample J6 had a considerably lower initial nitrogen concentration, explaining the lower fluorescence yield. Fluorescence decreases towards higher neutron irradiation times; note however that the ratio of centers in the negative charge state by trend increases for samples irradiated for \SI{> 50}{\hour} (though the variation between equally treated samples and within single samples is large). Cf. also Fig.~\ref{fig_nvm} for an estimate of the absolute \nvm concentration.

Electron-irradiated Type Ib HPHT samples generally exhibited fluorescence rates smaller than or at best equal to the neutron-irradiated samples. 

Lower energy electrons lead to a vastly better \nvratio ratio: while samples U5, U6, U7, U9 and U10 (\SI{6.5}{\MeV}) contain \SI{> 80}{\percent} \nvmnosp, samples S1, Q5 and Q6 (\SI{8}{\MeV}) contain \SI{< 30}{\percent}. The signal-to-noise ratio in our spectroscopic setup does not allow to quantify \nvm ratios lower than that upper bound. The total stopping power (average energy loss rate) for these two energies differs only by \SI{\approx 3}{\percent}, which cannot explain the discrepancy in the charge state ratio. The maximum transferrable energy in an elastic scattering event between an electron and a $\mathrm{^{12}C}$ nucleus however rises by a factor two from \SI{8.7}{\keV} at \SI{6.5}{\MeV} to \SI{12.8}{\keV} at \SI{8}{\MeV}\cite{campbell_radiation_2000}, potentially contributing to the different charge state distributions.

Annealing during irradiation has no clear influence on the \nvm abundance, it does however significantly improve the spin linewidth (samples U5, U6, U7, see Sec.~\ref{sec_odmr}).

For comparison, we also irradiated Type IIa CVD samples (initial $\mathrm{[N]\approx\SI{1}{ppm}}$) with neutrons and electrons. The resulting fluorescence intensities are lower by two to three orders of magnitude, as could be expected from the lower nitrogen content. The electron-irradiated samples were annealed during irradiation and show a maximum fluorescence at a dose of \SI{1e17}{\per \square \cm}, albeit at very low \nvm charge state ratios of \num{< 0.3}.

It is difficult to reliably calibrate the fluorescence rates observed in our experiments to absolute concentrations due to a large number of only inaccurately known parameters of the optical dynamics of \nvm and \nvz and of our experimental configuration. We tentatively estimate our best samples to contain \nvm concentrations in the range of several parts per million.

\subsection{Visible and UV absorption}
\label{sec_vis}
Upon visual inspection and in UV/Vis absorption spectra, the samples exhibit distinct characteristics depending on the treatment applied: Unirradiated Type Ib HPHT diamonds appear yellow due to the onset of absorption by photoionization of substitutional nitrogen (known as P1 or C centers) at \SI{500}{\nm} and their transparency at greater wavelengths (as illustrated by the trace for sample K3 in Fig.~\ref{fig_uvvis}). The absorption coefficient at \SI{400}{\nm} has been correlated with the substitutional nitrogen contents in Type Ib diamonds\cite{de_weerdt_determination_2008}. Unfortunately the sensitivity of our spectrometer was not sufficient to accurately resolve the high absorbance in this region, but we can place a lower bound of \SI{130}{ppm} on the $\mathrm{N_S}$ concentration in our raw Element-6 Type Ib samples.

\begin{figure*}[htb]
	\includegraphics[width=\textwidth]{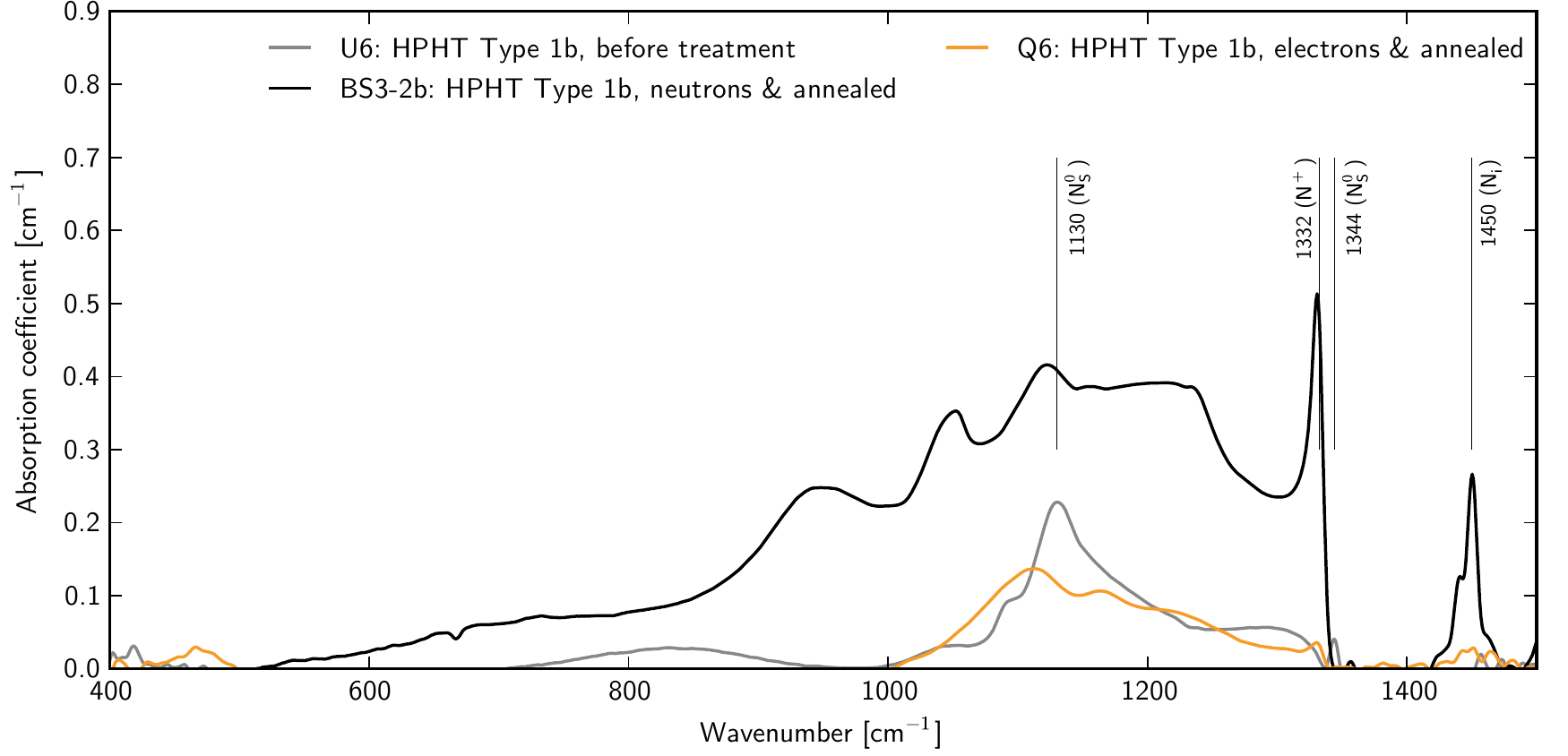}
	\caption{\label{fig_ftir}(Color online) FTIR absorption spectra four three HPHT Type Ib samples with different treatments. For a discussion of the indicated peaks, see the main text.}
\end{figure*}

Neutron-irradiated HPHT samples are opaque (at thicknesses of \SI{0.3}{\mm}) and black. Absorption sets in at \SI{\approx 700}{nm} due to excitation of phonon-excited states of the \nvm optical ground state (cf. trace BS3-2b in Fig.~\ref{fig_uvvis}). The \nvm ZPL is clearly visible as a peak in absorbance at \SI{637}{nm}, followed by the excitation phonon sideband (PSB) of that defect. Our samples reach an absorbance of 3 at around \SI{610}{\nm}, reaching the detection limit of our spectrometer. No further details could be discerned in the region towards smaller wavelengths.

The \nvm absorption spectrum is most clearly visible in samples that were irradiated with lower energy (\SI{6.5}{\MeV}) electrons and annealed during irradiation (trace U5 in Fig.~\ref{fig_uvvis}): the \nvm ZPL and one-phonon line (broad peak at \SI{\approx 610}{nm}) are discernible, the absorption PSB continues to dominate the absorption spectrum towards smaller wavelengths until \SI{480}{\nm}, where absorption due to photoionization of substitutional nitrogen sets in. A small peak in absorption is visible at \SI{389}{nm}, which is associated with a defect center created by radiation damage, possibly an interstitial nitrogen bound to a carbon atom\cite{zaitsev_optical_2001}. These samples show a deep purple hue. 

In samples that were irradiated with electrons of higher energy (\SI{8}{\MeV}) and not heated during irradiation, the same radiation damage feature is present, while otherwise their spectra differ quite clearly from the heated ones: The \nvm absorption spectrum is present but much weaker, instead the \nvz ZPL at \SI{575}{\nm} and the associated phonon sideband is much more pronounced, a fact reflected also in the fluorescence spectra presented in Sec.~\ref{sec_fluo}. Additionally, an absorption line appears at \SI{594.5}{\nm} which is associated with radiation damage in nitrogen-rich diamond and supposedly relates to a nitrogen+vacancy defect\cite{zaitsev_optical_2001}. Macroscopically, these samples appear orange. 

At wavelengths in the range \SIrange{700}{3000}{\nm}, all but the neutron-irradiated samples showed high transmission of \SI{\approx 60}{\percent} (data not shown). This figure is well accounted for by the Fresnel reflections off the two diamond surfaces, which amount to \SI{17}{\%} per surface at normal incidence. The neutron-irradiated samples showed increased absorption even at these long wavelengths (which are outside the NV absorption bands). The ZPL associated with the neutral vacancy center (GR1) -- a common irradiation damage product -- at \SI{741}{\nm} was not visible in any of the samples. The negative vacancy (ND1) absorption line at \SI{396}{\nm} was not visible either due to the strong absorption from substitutional nitrogen in all samples in this spectral region. 

\subsection{Infrared absorption}
\label{sec_ir}

Figure~\ref{fig_ftir} shows the FTIR absorption spectra of a number of diamond samples after subtracting the background from diffuse scattering and boxcar-averaging the data to filter out small etaloning oscillations. Untreated Type Ib diamond (cf. trace U6 in Fig.~\ref{fig_ftir}) primarily shows a broad vibrational absorption band at \SI{1130}{\per \cm} and a smaller peak at \SI{1344}{\per \cm}, both of which are typical for nitrogen-rich diamond and are associated with neutrally charged single substitutional nitrogen\cite{collins_anomaly_1982} (and hence labeled $\mathrm{N_{0}^{S}}$ in the plot).

After neutron irradiation, the spectrum is changed dramatically (cf. trace BS3-2b in Fig.~\ref{fig_ftir}): Several peaks arise that are associated with positively charged substitutional nitrogen ($\mathrm{N^{+}}$) \cite{lawson_existence_1998}, most prominently the sharp line at \SI{1332}{\per \cm}, but also the broader peaks at \SI{1045}{\per \cm} and \SI{945}{\per \cm}. The plateau connecting these peaks is supposedly due to absorption by bulk phonons which become allowed by the presence of defects breaking the translational symmetry of the lattice\cite{morelli_phonon_1993,smith_activation_1960}. While the peak indicating neutrally charged substitutional nitrogen at \SI{1344}{\per \cm} disappears, a pronounced peak at \SI{1450}{\per \cm} becomes apparent, which is ascribed to a local vibration of one nitrogen and two equivalent carbon atoms, with the nitrogen possibly occupying an interstitial position\cite{kiflawi_nitrogen_1996,woods_1450_1982} (labeled $\mathrm{N_i}$ in the plot). 

Samples irradiated with electrons (but not heated during irradiation; cf. trace Q6 in Fig.~\ref{fig_ftir}) show the \SI{1332}{\per \cm} line as well, but it is far less pronounced than the neutron-irradiated samples. Clear signals indicating the presence of neutrally charged substitutional nitrogen are absent, there are some weak peaks in the region related to interstitial nitrogen (\SI{1450}{\per \cm}). Otherwise the hump between \SI{1000}{\per \cm} and \SI{1332}{\per \cm} resembles the one seen in neutron-irradiated samples, albeit less pronounced, indicating radiation damage.

In summary, the difference between the unirradiated samples and the neutron-irradiated ones give a rather clear indication of the conversion process from neutral substitutional nitrogen to positively charged nitrogen plus \nvm centers (as seen in fluorescence): the nitrogen atoms not bound to a newly created vacancy serve as electron donors. In the samples irradiated with electrons at room temperature, this process takes place as well, but to a lesser extent: less positively charged nitrogen impurities emerge, and the ratio of \nvratio is much worse (i.e. more \nvz than \nvmnosp). Although the overall extent of lattice damage seems to be somewhat lower, the nature of that damage seems to prohibit the creation of the desired \nvm charge state. 

\begin{figure}[htb]
\includegraphics[width=\columnwidth]{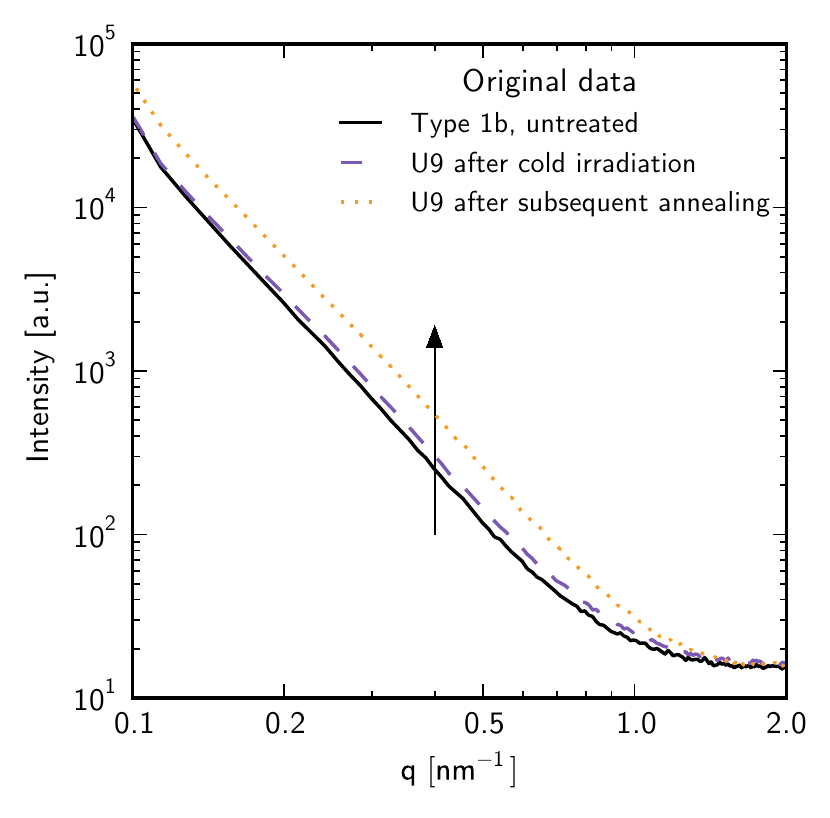}
\caption{\label{fig_saxs_a}(Color online) Small-angle X-ray scattering intensity vs. magnitude of the scattering vector for untreated diamond, diamond irradiated at room temperature before the final annealing step, and after annealing. The vertical arrow is a guide to the eye illustrating the increased scattering intensity after every treatment step.}
\end{figure}

\begin{figure}[htb]
\includegraphics[width=\columnwidth]{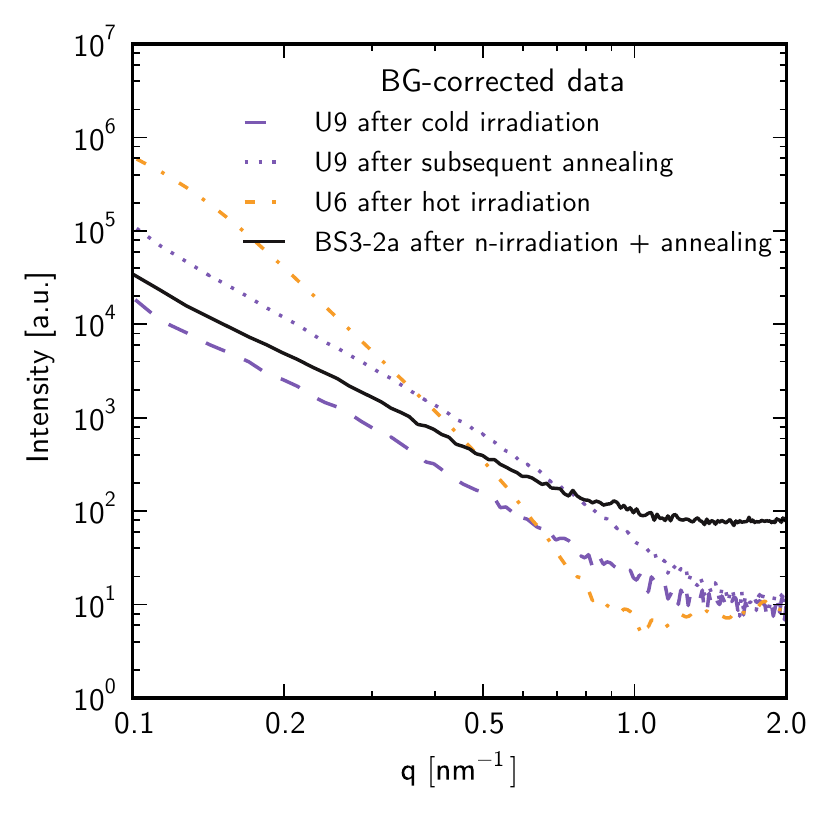}
\caption{\label{fig_saxs_b}(Color online) Background-corrected small-angle X-ray scattering intensity vs. magnitude of the scattering vector for a neutron-irradiated diamond sample, and two electron irradiated ones: sample U9 after cold irradiation, and after subsequent annealing, and sample U6 after hot irradiation.}
\end{figure}

\subsection{Small angle X-ray scattering}
\label{sec_saxs}
A number of diamond samples at different stages of treatment were investigated using SAXS in order to clarify the effect of the irradiation and annealing procedures on the structure of the crystal.

Fig.~\ref{fig_saxs_a} shows the original data for an untreated diamond in comparison with sample U9 (after irradiation with \SI{8}{\MeV} electrons at room temperature, i.e. ``cold irradiation''), both before and after annealing. In comparison to the untreated diamond, a slight but distinct intensity increase is visible. This can be attributed to the production of vacancies within the diamond by the high energy electrons. The formation of regions with different electron densities leads to an increased scattering contrast and scattering intensity.

During annealing the sample at \SI{900}{\celsius}, the vacancies become mobile \cite{twitchen_electronparamagneticresonance_1999}. They move towards the impurity platelet precipitations, which are visible as elongated streaks in the SAXS patterns. Platelet-like structures in diamond are always associated with nitrogen impurities\cite{moore_synchrotron_1993}. The migration process further enhances the scattering contrast and the scattering intensity, visualized by the arrow in Fig.~\ref{fig_saxs_a}.

This effect is more clearly visible in Fig.~\ref{fig_saxs_b}, which compares the background corrected data for sample U9 before and after annealing. Additionally, the scattering intensity curve for sample U6 is depicted, which was irradiated at \SI{\approx 800}{\celsius} (``hot irradiation''): it not only shows much higher scattering intensity, but also a shift of the scattering intensity distribution towards lower scattering vector magnitudes (i.e. larger structures in real space).

In order to quantify the effect of the different treatment methods, the background corrected scattering data were fitted with a function proposed by Beaucage\cite{Beaucage}.

\begin{align}
I(q) &= G \exp\left(-\frac{q^2 R^{2}_{g}}{3}\right)+ B \left[\frac{\text{erf}\left(\frac{q R_{g}}{\sqrt{6}}\right)^3}{q}\right]^{p} \nonumber \\
&+ \text{background}
\label{eqn:beaucage}
\end{align}

Here, $q$ is the magnitude of the scattering vector, $R_{g}$ is the radius of gyration, $\text{erf}()$ is the error function and B and G are material constants defined in Ref.~\onlinecite{Beaucage}. The fit parameters were $R_{g}$ and $p$. 

When comparing the fit results for the electron-irradiated samples, two tendencies can be observed: 

First, the radius of gyration increases from $R_{g}$=\SI{13}{\nm} to \SI{16}{\nm} and \SI{50}{nm} for the cold-irradiated sample before annealing, after annealing and the hot-irradiated sample, respectively. This increase correlates positively with the ratio of fluorescence from \nvm and \nvznosp: Luminescence from the desired charge state \nvm increases with size and intensity of the structural effects of the irradiation.

Second, the corresponding fractal dimension $p$ of these samples is 3.9, 3.6 and 4, respectively, where 4 is the theoretical value for a smooth surface between two phases. The value of 3.9 is within the error of the fit, and the decrease to 3.6 for the cold-irradiated sample after annealing can be interpreted as a roughening of the platelets formed by the nitrogen impurities (a so-called surface fractal). This roughening together with the slight increase of $R_{g}$ suggests the accretion of vacancies to the nitrogen platelets. 

For comparison with the electron-irradiated samples, we show data for the neutron-irradiated sample BS3-2a in Fig.~\ref{fig_saxs_b}. The main difference to the electron-irradiated samples is the strong scattering intensity at high q-values (in the range $q>\SI{1}{\per \nm}$). It is very probable that this originates from amorphous carbon, which causes more diffuse scattering in comparison to crystalline diamond. This is further confirmed by the change in the color of the diamond from transparent yellow to opaque black (see Sec.~\ref{sec_vis}).

\subsection{Optically detected magnetic resonance}
\label{sec_odmr}
\begin{figure}[tb]
\includegraphics[width=\columnwidth]{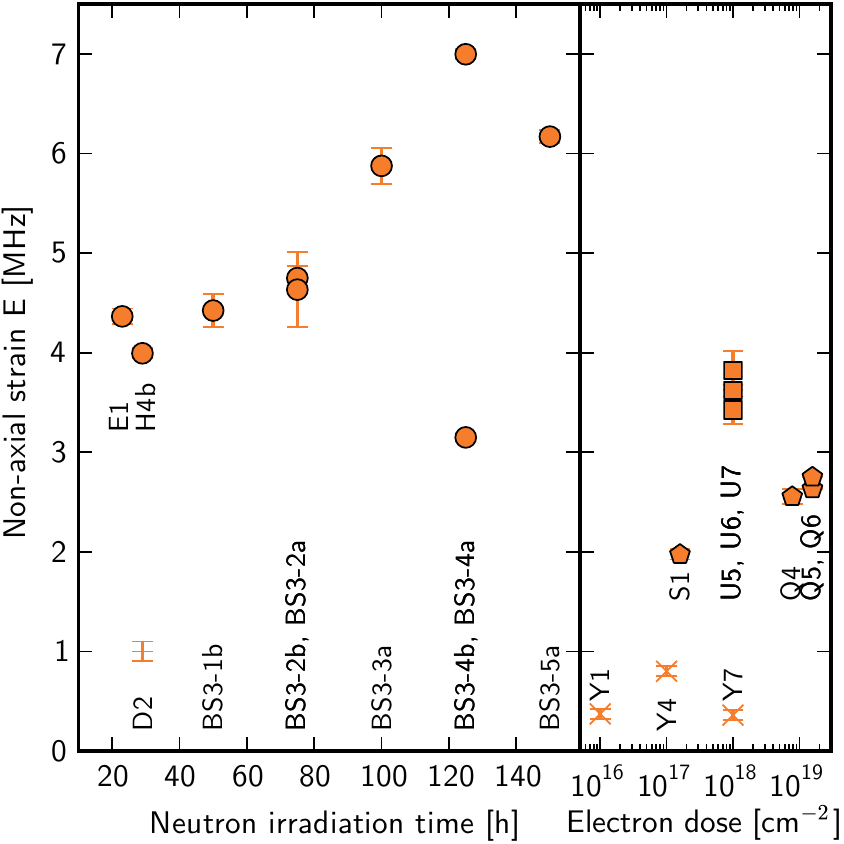}

\caption{\label{fig_E}Non-axial zero-field splitting parameter $E$ (strain) vs. neutron irradiation time or electron dose. Error bars indicate standard errors. For marker legend, see caption of Fig.~\ref{fig_fluo}.}
\end{figure}

\begin{figure}[tb]
\includegraphics[width=\columnwidth]{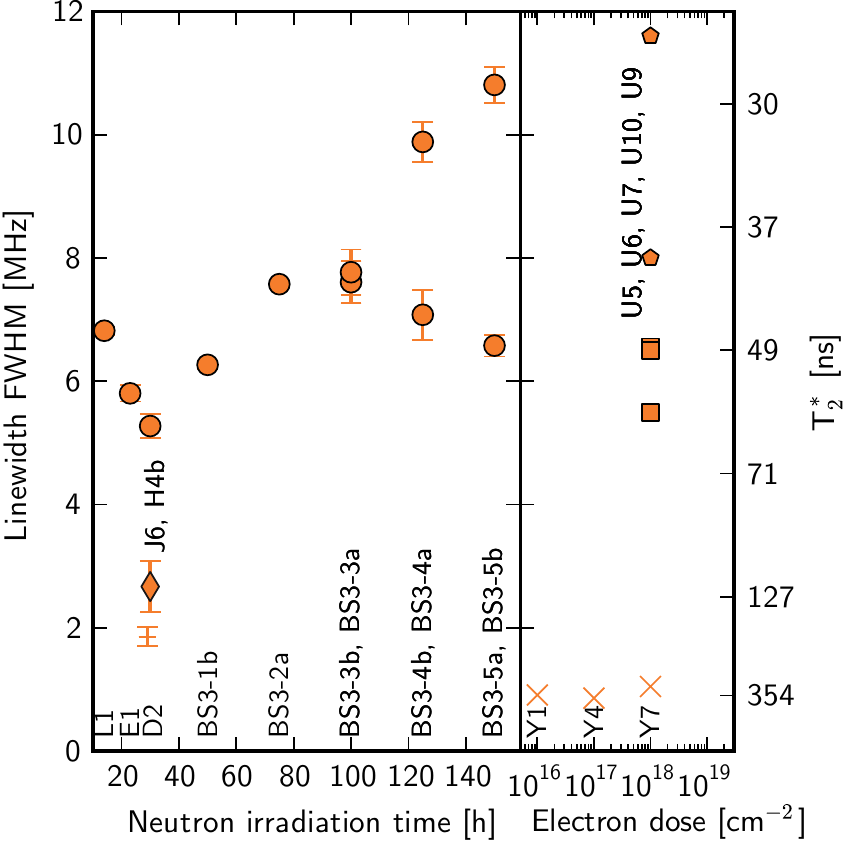}
\caption{\label{fig_width}Inhomogeneous linewidth $\Gamma_{\mathrm{FWHM}}$ vs. neutron irradiation time or electron dose. Error bars indicate standard errors. For marker legend, see caption of Fig.~\ref{fig_fluo}.}
\end{figure}

By simultaneously exciting at \SI{532}{\nm}, scanning the frequency of a microwave (MW) field across the \nvm zero-field splitting frequency $\mathrm{D=\SI{2.87}{\GHz}}$ and recording fluorescence intensity in the wavelength range \SI{> 645}{\nm}, the spin resonance properties of the \nvm defects in all samples were determined. The spin-dependent intersystem crossing rate between the \nvm excited triplet state $\mathrm{^3E}$ and a metastable dark singlet state leads to spin-dependent fluorescence rates and optically induced spin polarization in the \mszero\ spin state\cite{manson_nitrogen-vacancy_2006}. When a MW field drives the magnetic dipole transition of the spin, its population in \msone\ increases, leading to increased probability for inter-system crossing and a reduced fluorescence rate. Hence, the spin resonance spectra of \nvm can be read out optically. The spin-resonance properties of \nvmnosp\cite{loubser_electron_1978,he_paramagnetic_1993} and the details of the optically-detected magnetic resonance (ODMR) technique have been described elsewhere\cite{oort_optically_1988,jelezko_single_2006}.

The effective Hamiltonian of the \nvm ground state electron spin triplet coupled to the $\mathrm{^{14}N}$ nuclear spin triplet can be written as:

\begin{align}
\hat{H} = & D \left(S_z^2 - \frac{1}{3} \vec{S}^2\right) + E(S_x^2-S_y^2) \nonumber\\
& + A_{\perp}(S_x I_x + S_y I_y) + A_{\parallel} S_z I_z + \frac{1}{2} P \left(3 I_z^2 - \vec{I}^2\right) \nonumber\\
& + \mu_B g \vec{B} \vec{S},
\end{align}
where $D=\SI{2.87}{\GHz}$ is the zero-field splitting, $E$ is the non-axial strain, $A_{\parallel}=\SI{-2.3}{\MHz}$ and $A_{\perp}=\SI{-2.1}{Mhz}$ are the hyperfine coupling parameters, and $P=\SI{-5.0}{\MHz}$ is the nuclear quadrupole interaction parameter\cite{he_paramagnetic_1993}. $S$ and $I$ are used to denote the spin-1 operators for the electron and nuclear spin, respectively, and their components. The last term of the equation gives the electron Zeeman interaction, where $\mu_B=\SI{14}{\MHz \per \milli \tesla}$ is the Bohr magneton, $g=\num{2.003}$ is the \nvm g-factor and $\vec{B}$ is the external magnetic field. We neglect couplings to other nuclear spins, the Stark and nuclear Zeeman effects. 

In zero external field, this Hamiltonian results in four non-degenerate magnetic dipole transitions symmetric around $D$\cite{acosta_erratum_2011}, the inner two of which are separated by $\Delta\nu = 2 E$, the outer by $\Delta\nu = 2\sqrt{A_\parallel^2+E^2}$. We fit four Gaussians to the ODMR spin resonance data recorded in zero magnetic field (without compensating for small stray fields) in order to extract the parameters $D$ and $E$. $D$ (not shown) by trend decreases with increasing neutron irradiation time, by \SI{\approx 1}{\MHz} over the range of fluences we tested. The dependence of the non-axial strain $E$ on irradiation doses is shown in Fig.~\ref{fig_E}: for Type Ib HPHT samples it increases with increasing neutron irradiation time from \SI{4}{\MHz} to \SI{7}{\MHz}, indicating increasingly strong lattice damage. HPHT samples irradiated with electrons at room temperature show values of $E$ between 2 and \SI{3}{\MHz}, while those annealed during irradiation have higher strain, 3 to \SI{4}{MHz}. For the neutron-irradiated CVD sample, strain is \SI{\approx 1}{\MHz}. CVD samples irradiated with electrons and annealed simultaneously, we find $E$ to be between 360 and \SI{800}{kHz}. Note that all the values quoted here represent the most likely values of the distributions of $E$ across a large number of \nvm centers, which we assume to be two Maxwell-Boltzmann distributions extending to both sides from the ZFS value. Since several transitions overlap in zero external field, the shape of the distribution is hard to extract from our data.

In many application scenarios, the inhomogeneous dephasing time $T_2^*$ (also known as free induction decay time) of an \nvm spin ensemble is a critical parameter: it limits the sensitivity of magnetometers\cite{taylor_high-sensitivity_2008}, the coupling strength required to reach the strong coupling regime in cavity QED systems\cite{sandner_strong_2012}, and the number of modes that can be stored in a collective state quantum memory\cite{julsgaard_quantum_2013}. The inhomogeneous dephasing time of the spin ensemble is related to the FWHM spin resonance linewidth via $\Gamma = 1/(\pi T_2^*)$. Static differences in $D$ and $E$ between individual spins in an ensemble contribute to $\Gamma$, as well as inhomogeneous dipolar couplings to the surrounding spin bath. At low concentrations of paramagnetic impurities the naturally present $\mathrm{^{13}C}$ nuclear spins may be the dominant bath, while in typical Type Ib samples coupling to non-converted substitutional nitrogen defects dominates. Only at N to \nvm conversion efficiencies greater than \SI{\approx 50}{\%}, the interaction of the \nvm center spins among themselves constitute the limiting dephasing bath\cite{acosta_diamonds_2009}.

In order to characterize the inhomogeneous linewidths in our samples, we recorded ODMR spectra in a small external magnetic field which separates the resonance lines of the four crystallographically equivalent species of \nvm and lifts the degeneracy of the $m_S = 0$ to $m_S = \pm 1$ transitions. We chose the field strength such that the resulting Zeeman shift is considerably greater than the non-axial strain splitting while ensuring that the level mixing is still small. Thus, the effect of non-axial strain on width and position of the spin resonance lines goes to zero; axial strain and the spin bath remain as influencing factors. For each sample, we recorded one electron spin resonance line at a series of microwave powers and fitted each resonance dip with three Lorentzians of equal width and amplitude, with their separation fixed to the hyperfine splitting. We extrapolated the resulting linewidths to zero power in order to obtain the bare linewidths plotted in Fig.~\ref{fig_width}.

For a concentration of positively charged substitutional nitrogen on the order of \SI{200}{ppm}, the resulting contribution to \nvm linewidth can be estimated to be \SI{\approx 4}{\MHz}. In all our neutron-irradiated HPHT samples with that initial amount of nitrogen, the linewidth is larger: at the lower neutron doses, it is between \num{5.2} and \SI{6.8}{MHz} and tends to increase (albeit with a huge variation between some of the samples) with higher doses, reaching \SI{10.8}{\MHz}. This indicates that variations in lattice strain and/or other paramagnetic impurities created by irradiation limit the linewidth. 

In samples irradiated with electrons and annealed simultaneously, the linewidth is comparable to the values found at lower neutron doses (see samples U5, U6, U7, indicated as squares in Fig.~\ref{fig_width}), and lower than in samples that were kept at room temperature during irradiation (U9, U10, pentagons). We presume that annealing during irradiation assists the healing of crystal damage before stable vacancy clusters -- and hence excess strain -- can form.

In all the Type Ib samples, the hyperfine structure cannot be resolved -- with the exception of sample J6, which is a Sumitomo HPHT stone with lower initial nitrogen concentration (\SI{100}{ppm}): here, the linewidth observed (\SI{2.6}{MHz}) is compatible with the nitrogen-limited spin bath contribution and only slightly larger than the hyperfine splitting (\SI{2.3}{\MHz}), allowing to resolve dips between the hyperfine transitions. 

\begin{figure}[tb]
\includegraphics[width=\columnwidth]{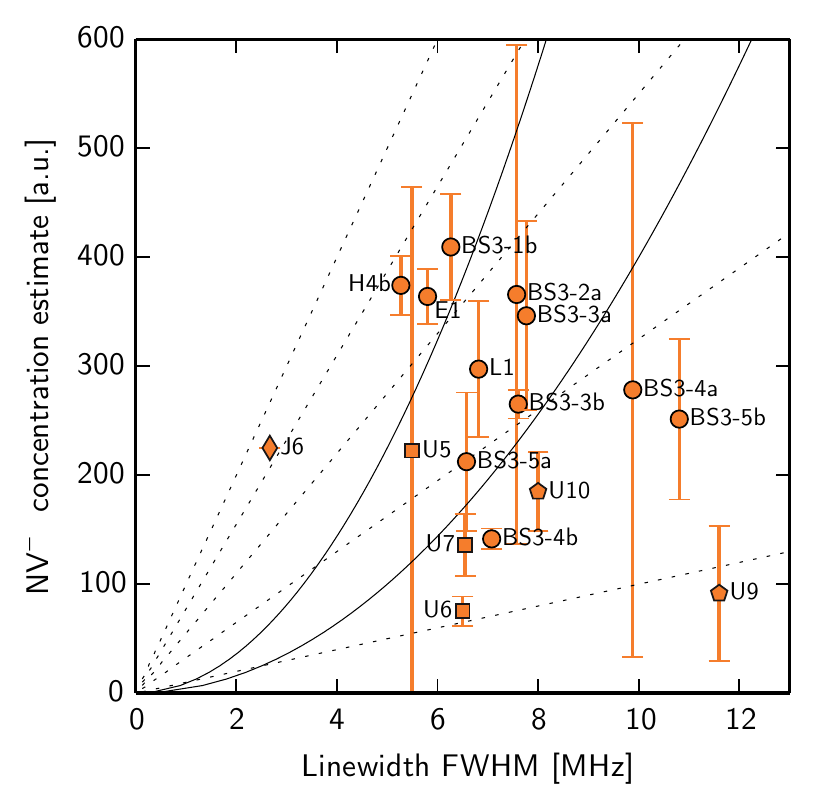}
\caption{\label{fig_nvm}Estimated relative [\nvmnosp] concentrations vs. spin resonance linewidths. Dashed lines indicate contours of the figure of merit (FOM) relevant for hybrid quantum systems $\mathrm{FOM_{HQS}} \sim \sqrt{[\mathrm{NV^{\mbox{-}}}]} / \Gamma$. Solid lines are countours of the FOM relevant for magnetometry, $\mathrm{FOM_{magn}} \sim \sqrt{[\mathrm{NV^{\mbox{-}}}] / \Gamma}$, where $\Gamma$ is the linewidth. The concentration is estimated from the total fluorescence intensity and the \nvratio ZPL intensity ratio and takes into account the different fluorescence spectra, excitation efficiencies, Huang-Rhys factors and fluorescence yields. For marker definitions see Fig.~\ref{fig_fluo}.}
\end{figure}

For comparison, we also irradiated Type IIa CVD samples with neutrons and electrons. In this class of material, $\mathrm{^{13}C}$ nuclear spins (at a natural abundance of \SI{1.1}{\%}) are the dominant spin bath. Experiments with single \nvm centers allow to evaluate the contribution of these nuclear magnetic moments to the linewidth, with results ranging from \num{0.2} to \SI{1}{MHz}\cite{mizuochi_coherence_2009,dutt_quantum_2007}. Our neutron-irradiated sample has a linewidth of \SI{1.8}{MHz}, which probably results from excess lattice strain (the sample was subject to an overly high neutron dose, given its nitrogen content). For the electron-irradiated samples (annealed during irradiation) we obtained better values ranging from \SI{850}{\kHz} to \SI{1}{\MHz}.

\section{Discussion \& Conclusion}
\label{sec_discussion}
Both magnetometry and hybrid quantum systems applications require high \nvm density while maintaining a low linewidth (i.e. high dephasing time). For an electron spin-based magnetometer, the minimum detectable magnetic field $\delta\! B$ (limited by spin-projection noise) is given by~\cite{acosta_diamonds_2009,budker_optical_2007}
\begin{align}
\delta\! B \sim \frac{1}{g_S \mu_B} \frac{1}{R \sqrt{\eta}} \frac{1}{\sqrt{N t T_2^*}},
\end{align}
where $R$ is the measurement contrast, $\eta$ is the detection efficiency, $N$ is the number of spin centers, and $t$ is the integration time. Hence, the primary figure of merit (FOM) to optimize for this application is $\mathrm{FOM_{magn}} \propto \sqrt{[\mathrm{NV^{\mbox{-}}}] / \Gamma}$. Recent results~\cite{kim_electron_2012} however indicate that the measurement contrast decreases systematically with increasing irradiation dose, limiting the practically achievable sensitivity in spite of an increasing \nvm density. The work cited tentatively ascribes the reduction of contrast to a negative effect of increased irradiation damage on optical polarization efficiency and/or the spin-dependence of the fluorescence rate. Although we observe rather low ODMR contrast in all of our samples on the order of \num{0.1}-\SI{0.5}{\percent} at the lowest MW powers used, we cannot systematically verify this observation from our data due to differing experimental parameters.

When forming a hybrid quantum system (HQS) by coupling a mode of the electromagnetic field inside a cavity to collective excitations in an ensemble of two-level systems\cite{schuster_high-cooperativity_2010,kubo_strong_2010,amsuss_cavity_2011}, the coupling strength $g$ scales with the square root of the density of emitters within the mode volume of the cavity. At the same time, a large dephasing time is desirable in order to enable a large number of coherent operations. Larger inhomogeneous broadening however leads to a decrease of the ratio $\frac{g}{\Gamma}$, which is required to be $>1$ for strong coupling\cite{sandner_strong_2012}. Hence, we regard $\mathrm{FOM_{HQS}} \propto \sqrt{[\mathrm{NV^{\mbox{-}}}]} / \Gamma$ as the FOM relevant for this case. Figure~\ref{fig_nvm} gives an overview of how our samples perform according to these two FOM, the contour lines of which are indicated in the plot. 

In summary, for experiments not requiring optical transparency, we find optimal parameters for neutron-irradiated Type Ib diamonds which received rather low fluences on the order of \SI{1e17}{\per \square \cm}. Sample J6, the only HPHT sample in this study manufactured by Sumitomo shows a smaller \nvm density -- which was to be expected from its lower initial nitrogen concentration -- stands out due to its substantially lower linewidth. We suspect that this is due to a more favorable material composition (strain, impurities). More experiments with this material will be necessary to confirm this observation. Electron irradiation is advantageous where the samples are required to have high optical transparency. Irradiating with lower energy electrons greatly increases the \nvm ratio, while annealing during irradiation decreases the resulting inhomogeneous broadening and further improves the optical quality. We presume these improvements to be due to more effective lattice healing processes compared to annealing after irradiation.

\begin{acknowledgments}
We are indebted to Fedor Jelezko for many fruitful discussions. We thank Hinrich Grothe at the Institute for Material Chemistry, TU Vienna, for granting us access to his spectrophotometers.  We are grateful to Klaudia Hradil at the X-ray service center of TU Vienna for additional X-ray diffraction measurements. We greatly appreciate assistance with neutron irradiation from the reactor and radiation safety teams at Atominsitut. We thank staff at the Istituto per la Sintesi Organica e la Fotoreattivit\`{a} in Bologna, Italy, for carrying out the electron irradiation. T. N. and K. B. acknowledge the Vienna Graduate School for Complex Quantum Systems (CoQuS), and S. P. the Doctoral Program ``Building Solids for Function - Solids4Fun'', both funded by the Austrian Science Fund FWF which also provided financial support to J. A. and H. P. (project no. I449) as well as J.S. (Wittgenstein Award). This research was funded by the EU FP7 collaborative project DIAMANT and by the Austrian Research Promotion Agency FFG via the project NAP/PLATON.
\end{acknowledgments}

\bibliography{neutron_irradiation_paper}

\end{document}